\documentclass[twocolumn,showpacs,preprintnumbers,amsmath,amssymb,aps,pra]{revtex4}
\usepackage{latexsym}
\usepackage{dcolumn}
\usepackage[english]{babel}
\usepackage{bm}
\usepackage{graphicx}
\input epsf
\begin{document}
\title{Two-Photon Doppler cooling of alkaline-earth-metal and ytterbium atoms}

\author{Wictor C. Magno, Reinaldo L. Cavasso Filho and Flavio C. Cruz}
\email[]{flavio@ifi.unicamp.br}
\affiliation{Instituto de F\'{i}sica 'Gleb Wataghin', Universidade Estadual de Campinas CP.6165, Campinas, SP, Brazil}


\begin{abstract}
A new possibility of laser cooling of alkaline-earth-metal and Ytterbium atoms using a two-photon transition is analyzed. We consider a $^1S_0$ - $^1S_0$ transition, with excitation in near resonance with the $^1P_1$ level. This greatly increases the two-photon transition rate, allowing an effective transfer of momentum. The experimental implementation of this technique is discussed and we show that for Calcium, for example, two-photon cooling can be used to achieve a Doppler limit of 123 microKelvin.~The efficiency of this cooling scheme and the main loss mechanisms are analyzed.

\end{abstract}


\maketitle

\section{Introduction}

Laser cooling of neutral atoms and ions has played a fundamental role in many fields as metrology, atom and quantum optics and Bose-Einstein condensation. For a strong and closed single photon atomic transition (linewidth $\Delta \nu$ of a few MHz), a high scattering rate can quickly reduce the atomic temperature (or velocity) down to a ``Doppler" limit given by the spontaneous emission rate, typically in the milliKelvin range. The multilevel atomic structure is essential in other cooling techniques, as polarization gradient \cite{Dalibard89}, \cite{Weiss89}, which allow the achievement of microKelvin temperatures. Although these techniques have been widely applied in metal-alkaline elements, they are not applicable to the alkaline-earth and Ytterbium atoms because of their simpler level structure, with non-degenerate ground states and no hyperfine structure. Thus, until recently the smaller temperatures achieved with these elements were in the milliKelvin range, given by the Doppler limit of the resonant $^1S_0$ - $^1P_1$ cooling transition. MicroKelvin temperatures, close to the recoil limit, have been achieved for strontium, by Doppler cooling using the narrower $^1S_0$ - $^3P_1$ intercombination transition ($\gamma/2\pi = 7.6~kHz$) \cite{Katori99} and also for Ytterbium ($\gamma/2\pi = 182~kHz$) \cite{Kuwamoto99}, an alkaline-earth-like atom. Cooling using the intercombination transition has also been demonstrated for Calcium, where the relatively long lifetime of the $^3P_1$ level ($\tau = 0.38~ms$, $\gamma/2\pi = 408 Hz$) had to be reduced by coupling it to another level \cite{Curtis01}, \cite{Binnewies01} in order to make the cooling process effective. This ``quenching'' cooling scheme has also been used several years ago to cool mercury ions to the zero-point energy of a trap \cite{Jim89}. With the exception of Ytterbium, cooling on the intercombination transition was used as a second stage, after initial pre-cooling to milliKelvin temperatures with the strong $^1S_0$ - $^{1}P_1$ dipole transition. Although this narrow line cooling can reach temperatures near the recoil limit, only a small fraction of the atoms cooled by the strong $^1S_0$ - $^1P_1$ line can be transfered to the microKelvin regime. For Calcium, for example, only about $15\%$ of the atoms captured in a conventional MOT were transferred to lower temperatures by ``quenching'' cooling \cite{Curtis01}, \cite{Binnewies01}. Another possibility for obtaining cold Calcium atoms was recently demonstrated by optically pumping a fraction of atoms captured in a conventional MOT into the metastable $^3P_2$ state, and application of a second stage of laser cooling on the $^3P_2$ - $^{3}D_3$ transition to produce metastable Calcium at microKelvin temperatures \cite{Andreas02}.

In this paper we present a new scheme for laser cooling of alkaline-earth and Ytterbium
atoms using a two-photon $^1S_0$ - $^1S_0$ transition. Although laser cooling involving 
two photons takes place for example in Raman cooling \cite{Kasevich92} and ``quenching'' 
cooling, so far no two-photon Doppler cooling has been demonstrated. We show that two-
photon cooling can be quite efficient and might be used as a second cooling stage, after pre-cooling with the conventional method. The experimental setup should be much 
simpler than in ``quenching'' cooling, requiring only one additional laser, with linewidth on the order of a MegaHertz. In addition, a significant temperature reduction, with respect to the one-photon Doppler limit, can be achieved. In Section II, we describe the feasibility of two-photon cooling and present the calculations of the transition rates for the one-photon $^1S_0$ - $^1P_1$ and two-photon $^1S_0$ - $^1S_0$ cooling transitions. In Section III, we present an analysis of the temperature limit that can be achieved by this new cooling scheme. Section IV discuss the main loss mechanisms, considering $^{40}Ca$ as an example. Finally, the conclusions are summarized in Section V.

\section{Two-Photon Transition Rates}

Let us consider the interaction of an atom with two copropagating laser beams, with wavenumbers $k_{1} = 2\pi/\lambda_{1}$ and $k_{2} = 2\pi/\lambda_{2}$ and frequencies $\omega_{1}$ and $\omega_{2}$. Figure 1 (a) shows a diagram of the relevant level structure of alkaline-earth and Ytterbium atoms. The ground state is represented by $\left|g \right\rangle$ = $(ns^{2})^{1}S_{0}$ and the excited states are $\left|r \right\rangle$ = ($ns np)^{1}P_{1}$, and $\left|e \right\rangle$ = $[ns (n+1)s]$ $^{1}S_{0}$. From level $\left|e \right\rangle$, the atoms quickly decay to the ground state by spontaneous emission via the intermediate state $\left|r \right\rangle$, with relaxation rates $\gamma_{2}$ and $\gamma_{1}$. Table I presents some relevant parameters for the most abundant isotopes of $Mg$, $Ca$, $Sr$ and $Yb$. 

\begin{center}
\begin{tabular}{p{1.1in} p{.5in} p{.5in} p{.5in} p{.5in} p{.5in}} 
\hline
\hline
~Parameter & $^{24}Mg$ & $^{40}Ca$ & $^{88}Sr$ & $^{174}Yb$ \\ \hline
$\lambda_{1}(nm)$ &~285.2 &~422.8 &~460.7 &~398.8 \\ 
$\gamma_{1}/2\pi(MHz)$ &~80.95 &~34.63 &~~4.48 &~28.01 \\ 
$I_{1s}(mW/cm^2)$ &~456.0 &~~59.9 &~~~6.0 &~~57.7 \\ 
$\lambda_{2}(nm)$ &1182.8 &~1034.4 &~1130 &1077.3 \\ 
$\gamma_{2}/2\pi(MHz)$ &~~4.14 &~~4.77 &~~2.96 &~~4.81 \\ 
$I_{2s}(mW/cm^2)$ &~~~0.3 &~~~0.6 &~~~0.3 &~~~0.5 \\ 
$v_{r} (cm/s)$ &~~7.14 &~~3.31 &~~1.39 &~~0.79 \\ 
$T_{r} (\mu K)$ &~14.92 &~~5.30 &~~2.04 &~~1.31 \\ \hline
\hline
\end{tabular}
\end{center}

\begin{center}
Table I. Parameters of interest for one- and two-photon cooling in some alkaline-earth-metal and Yb atoms.
\end{center}

We see that for all these atoms the first laser is in the blue or ultraviolet region, while the other laser is at the near infrared. The resonant saturation intensity for the blue transition ($^1S_{0}$ - $^1P_{1}$) is $I_{1s}$, while $I_{2s}$ is the one for the infrared transition ($^1P_{1}$ - $^1S_{0}$). The recoil velocity, after two-photon absorption, is given by $v_{r} = \hbar (k_{1} + k_{2})/M$, where $M$ is the atomic mass. The recoil temperature is then $T_{r} = M v_{r}^2/k_{B}$, where $k_{B}$ is the Boltzmann constant. Both are obviously higher than in the case of the single photon $^1S_0$ - $^{1}P_1$ cooling transition.

Two-photon transition rates are dependent on the light beam intensities and detunings from real levels and usually are much smaller than single-photon dipole allowed transition ones. Nevertheless, they can be quite strong for spectroscopic purposes, allowing the implementation of powerful Doppler-free techniques \cite{Demtroder96}, widely used over the last thirty years. However, in order to reduce the atomic velocity, they must be high enough to allow an effective and fast transfer of momentum from the light fields to the atom. The general expression for the two-photon transition rate between levels $\left|g \right\rangle$ and $\left|e \right\rangle$ is given by \cite{Cagnac73} and \cite{Loudon83}:

\begin{eqnarray}
\Gamma_{ge} = \left| \sum_{r} \frac {\left\langle e \left|H_{2}\right|r\right\rangle \cdot \left\langle r \left|H_{1}\right|g\right\rangle}{\hbar\Delta_{r} - i\left( \hbar\gamma_{_1}/2 \right)} \right|^2 \cdot \frac{\gamma_{_2}}{\left[\Delta_{_2}^2 + \left(\gamma_{_2}/2 \right)^2 \right]},
\label{eq:Gge1} 
\end{eqnarray}

\bigskip

\noindent
where the matrix elements of the interaction Hamiltonian between the atomic levels are  $\left\langle e \left|H_{2}\right|r\right\rangle$ and $\left\langle r \left|H_{1}\right|g\right\rangle$. The detunings, including the Doppler shifts, are $\Delta_{r} = \omega_{gr} - \omega_{_1} + k_{_1} v$, and $\Delta_{2} = \omega_{eg} - \omega_{_1} - \omega_{_2} + (k_{_1} + k_{_2}) v$, with the Bohr frequencies between the atomic levels $\left|l \right\rangle$ and $\left|m \right\rangle$ represented by $\omega_{lm}$. The laser detuning relative to level $\left|r \right\rangle$ is defined as $\delta \omega_{_1} = \omega_{_1} - \omega_{gr}$. The first term of the expression involves a sum on the intermediary levels $\left|r \right\rangle$, which are coupled to $\left|g \right\rangle$ by allowed single-photon transitions. The second factor gives the spectral line profile of the two-photon transition of a single atom, which has a two-photon detuning $\delta \omega_{_2} = \omega_{_1} + \omega_{_2} - \omega_{eg}$. In order to preserve angular momentum in a $\Delta J = 0$ transition ($^1S_0$ - $^1S_0$), we are also assuming $\sigma_{1}(-)$ and $\sigma_{2}(+)$ polarizations for the two laser beams at $\lambda_{1}$ and $\lambda_{2}$ respectively (see Fig.1~(b)). Although $\sigma_{1}(+)$ and $\sigma_{2}(-)$ photons can be absorbed from opposite directions, this possibility will not be considered here. Polarization selection rules for two-photon transitions are discussed in \cite{Demtroder96} and \cite{Bonin84}. For alkaline-earth-metals and Ytterbium, Eq.~(\ref{eq:Gge1}) can be written as:

\begin{eqnarray}
\Gamma_{ge} = \frac{4 \cdot S_{1} \cdot S_{2} \cdot \gamma_{_2}}{\left[1 + S_{1} + \left(2 \Delta_{_r}/\gamma_{_1}\right)^2 \right] \cdot \left[1 + S_{1} \cdot S_{2} + \left(2 \Delta_{_2}/\gamma_{_2}\right)^2 \right]},
\label{eq:Gge2}
\end{eqnarray}

\bigskip

\noindent
where the sum over intermediary levels $\left|r \right\rangle$ is reduced to only one term, involving the real level $^1P_{1}$. We used in this equation the saturation parameters $S_{1} = 2\left| \Omega_{1} \right|^2/\gamma_{_1}^2 = I_{1} (\omega_{_1})/I_{1s}$, and $S_{2} = 2\left| \Omega_{2} \right|^2/\gamma_{_2}^2 = I_{2} (\omega_{_2})/I_{2s}$, where $\Omega_{j} = - \mu_{j} E_{j}(\omega_{_j}) /\hbar$ $(j = 1, 2)$ are the Rabi frequencies associated with the applied fields $E_{j}(\omega_{_j})$, and the dipole operators $\mu_{j}$. We have introduced in this last equation the possibility of saturation of the one- and two-photon transitions.

The transition rate, $\Gamma_{gr}$, for the strong one-photon $^1S_0$ - $^1P_1$ dipole transition is given by \cite{Metcalf99}:

\begin{eqnarray}
\Gamma_{gr} = \frac{S_{1} \cdot (\gamma_{_1}/2)}{\left[1 + S_{1} + \left(2 \Delta_{_r}/\gamma_{_1}\right)^2 \right]},
\label{eq:Ggr}
\end{eqnarray}

\bigskip

Figures 2(a) and 2(b) show the transition rates $\Gamma_{ge}/2\pi$ and $\Gamma_{gr}/2\pi$, as a function of the one-photon detuning $\left|\Delta_{r}\right|$, normalized by $\gamma_{_1}$, for a two-photon detuning $\left|\Delta_{2}\right| = \gamma_{_2}/2$, and saturation parameters $S_{1} = 0.1$ and $S_{2} = 3$. The calculations were made for Magnesium (dashed), Calcium (solid), Strontium (dash-dot) and Ytterbium (doted curve). 
The chosen saturation parameters made the one- and two-photon transition rates of the same 
magnitude. For $^{40}Ca$, for example, they correspond to (see Table I) $I_{1}(\omega_{1}) \approx 6 mW/cm^2$, and $I_{2}(\omega_{2}) \approx 2 mW/cm^2$. In fact over 100 mW of laser radiation at 422.8 nm can be generated by frequency doubling near infrared diode lasers \cite{Manoel02} or a Ti:sapphire laser \cite{Onisto02}. Radiation at 1034 nm can be generated with a Ti:sapphire laser or with a Yb:YAG laser. With this last option, powers over 500 mW are readily achieved. Even at low power levels the two-photon transition rate can still be significant. We can see in Fig. 2~(a) that, for $\left|\Delta_{_r}\right| < \gamma_{1}$, the two-photon transition rate is about 1~MHz, therefore a little smaller than the rates of the conventional cooling transitions of metal-alkaline atoms. If we assume, for example, red detuning of the incident lasers, $\Delta_{_r} = - \gamma_{_1}/2$ and $\Delta_{_2} = - \gamma_{_2}/2$, the two-photon transition rate will be $\Gamma_{ge}/2\pi = 1.2~MHz$ for Calcium. For these parameters, the single-photon transition rate is $\Gamma_{gr}/2\pi \ = 826~kHz$.

The two-photon transition rate can be much higher than the one-photon rate, if  the  saturation parameter $S_{2}$ is much higher than $S_{1}$. However, as we will see in the next section, increasing the second laser intensity results in an increase of the equilibrium temperature, due to the heating caused by spontaneous emission.

\section{Doppler - cooling Limit}

The temperature limit in laser cooling is determined by a balance between damping forces and heating due to spontaneous emission. For a two-level system, considering an one-photon process this Doppler limit is $k_{B}T_{D} = \hbar \gamma_{1}/2$, while the recoil limit is given by $k_{B}T_{r} = \hbar^2 k_{1}^2/M$. In general, $T_{D} >> T_{r}$, for one-photon cooling transitions of alkalis.

\bigskip

\noindent
\textbf{A. Two-photon Process}

\bigskip

The Doppler temperature limit can be estimated by the balance between heating and cooling, giving $k_{B}T_{D} = D/\alpha$ \cite{Metcalf99}, where $D$ is the diffusion constant and $\alpha$ is the damping coefficient. Lets consider one atom interacting with two copropagating $\sigma_{_2}(+)$, $\sigma_{_1}(-)$ laser beams (see Fig.1~(b)). The direct excitation from $\left|g \right\rangle$ to $\left|e \right\rangle$ by the simultaneous absorption of two photons, $\omega_{_1} + \omega_{_2}$,  occurs with a rate $\Gamma_{ge}$, given by Eq.~(\ref{eq:Gge2}). From the upper level $\left|e \right\rangle$, the atom spontaneously decays to the intermediate $\left|r \right\rangle$ level with a rate $\gamma_{_2}$ and, from this level, with a rate $\gamma_{_1}$ to the ground state. On average, the time it takes one atom to absorb simultaneously two photons and go back to the ground state by spontaneous cascade decay is given by:

\begin{eqnarray}
\Delta t = \Gamma_{ge}^{-1} + \gamma_{_2}^{-1} + \gamma_{_1}^{-1}.
\label{eq:Dt}
\end{eqnarray}

\bigskip

\noindent
If the atom is moving in the opposite direction of the laser beams (left side of Fig.~1-b),  then a mean radiation force, $F_{+}$, can be written as the ratio between the momentum change $\Delta p = \hbar (k_{1} + k_{2})$, and the time interval $\Delta t$:

\begin{widetext}
\begin{eqnarray} 
F_{+} = \frac{4\hbar (k_{1} + k_{2})S_{1}S_{2}\gamma_{_2}}{\left(1 + S_{1} + 4x_{_1}^2 \right) \cdot \left(1 + S_{1}S_{2} + 4x_{_2}^2 \right) + 4 S_{1}S_{2} \cdot (1 + \gamma_{_2}/\gamma_{_1})} \times \nonumber 
\\ [0.5cm] 
\left\{1 - \frac{8 \cdot \left(1 + S_{1}S_{2} + 4x_{_2}^2\right)\cdot (x_{_1}k_{_1}v/\gamma_{_1}) + 8 \cdot \left(1 + S_{1} + 4x_{_1}^2\right)\cdot \left[x_{_2} (k_{_1} +  k_{_2})v/\gamma_{_2}\right]}{\left(1 + S_{1} + 4x_{_1}^2\right) \cdot \left(1 + S_{1}S_{2} + 4x_{_2}^2\right) + 4 S_{1}S_{2} \cdot (1 + \gamma_{_2}/\gamma_{_1})} \right\},
\label{eq:Fr} 
\end{eqnarray}
\end{widetext}

\bigskip
\bigskip

\noindent
where $x_{_i}$ = $\delta\omega_{_i}/\gamma_{_i}~(i = 1, 2)$ are the detunings normalized by the natural linewidths of the transitions, and we have disregarded terms on the order of $(kv)^{2}/\gamma^2$. Adding one pair of $\sigma_{_1}(+)$ and $\sigma_{_2}(-)$ laser beams counterpropagating to the first ones (right side of Fig.~1-b), an additional force $F_{-}$  will act on the atom, and we have a configuration of an one-dimension two-photon optical molasses. In this configuration, the radiation pressure reduces to

\begin{eqnarray} 
F = F_{+} + F_{-} = - \alpha_{_2} \cdot v,
\label{eq:Ftotal} 
\end{eqnarray} 

\bigskip
\noindent
where the damping coefficient, $\alpha_{_2}$, is given by:

\begin{widetext}
\begin{eqnarray} 
\alpha_{_2} = \frac{- 64 \hbar (k_{_1} + k_{_2})S_{1}S_{2}\gamma_{_2}}{\left\{ \left(1 + 2S_{1} + 4x_{_1}^2 \right) \cdot \left(1 + 4S_{1}S_{2} + 4x_{_2}^2 \right) + 4 S_{1}S_{2} \cdot (1 + \gamma_{_2}/\gamma_{_1}) \right\}^2} \times \nonumber \\ [0.5cm]
\left\{(x_{_1}k_{_1}/\gamma_{_1}) \cdot \left[1 + 4S_{1}S_{2} + 4x_{_2}^2 \right] + \left[x_{_2}(k_{_1} + k_{_2})/\gamma_{_2}\right] \cdot \left[1 + 2S_{1} + 4x_{_1}^2\right] \right\},
\label{eq:alpha2} 
\end{eqnarray} 
\end{widetext}

\bigskip

\noindent
which leads to a kinetic energy losing rate $(dE/dt)_{cool} = F \cdot v =  - \alpha_{_2} \cdot v^2$. The damping coefficient is positive for red detunings and implies in a damping force for all velocities, if $\delta \omega_{_1}, \delta\omega_{_2} < 0$, similar to the case of ``one-photon optical molasses''. For low intensities and far from saturation of the transitions ($S_{1} << 1$, $S_{1} \cdot S_{2} << 1$) we have:

\begin{widetext}
\begin{eqnarray} 
\alpha_{_2} = \frac{- 64 \hbar (k_{_1} + k_{_2})S_{1}S_{2}\gamma_{_2}}{\left(1 + 4x_{_1}^2 \right) \cdot \left(1 + 4x_{_2}^2 \right)} \left\{ \frac{x_{_1}k_{_1}/\gamma_{_1}}{1 + 4x_{_1}^2} +  \frac{x_{_2}(k_{_1} + k_{_2})/\gamma_{_2}}{1 + 4 x_{_2}^2} \right\},
\label{eq:alpha2b} 
\end{eqnarray}
\end{widetext}

\bigskip

The momentum diffusion constant in a ``two-photon molasses'', $D_{2}$, can be estimated through a straightforward generalization of the analysis of the random walk process in ``one-photon molasses'' made by Lett \textit{et al.} \cite{Lett89}. Although the damping force reduces the average velocity of the atoms to zero, the fluctuations of this force produce heating, due the spontaneous emission, which leads to a spread in the mean square momentum:

\begin{eqnarray}
d(p^2)/dt = 2 \cdot \hbar^2 \cdot (k_{_1}^2 + k_{_2}^2) \cdot \Gamma_{ge} = 2 \cdot D_2.
\label{eq:dp2dt}
\end{eqnarray}

\bigskip

\noindent
In analogy to \cite{Lett89}, the resultant momentum diffusion constant in a ``two-photon molasses'' is:

\begin{eqnarray}
D_2 = \frac{4 \hbar^2 (k_{_1}^2 + k_{_2}^2) \cdot (2S_{1}) \cdot (2S_{2}) \cdot \gamma_{_2}}{\left(1 + 2S_1 + 4x_{_1}^2 \right) \cdot \left(1 + 4 S_1S_2 + 4x_{_2}^2\right)},
\label{eq:D2}
\end{eqnarray}

\bigskip

\noindent
where the saturation of the one- and two-photon transitions by the two pairs of counter-propagating beams was taken into account, and we have disregarded the Doppler shifts: $\left|v\right| << \gamma_{_1}/k_{_1}$, $\gamma_{_2}/(k_{_1} + k_{_2})$.

The rate of increase of kinetic energy in this ``two-photon molasses'' is given by $(dE/dt)_{heat} = d(p^2/2M)/dt = D_{2}/M$. When the equilibrium is reached, the heating and cooling rates are equal, that is, $(dE/dt)_{cool} + (dE/dt)_{heat} = 0$. Therefore, using equations (\ref{eq:alpha2b}) and (\ref{eq:D2}), the Doppler temperature is estimated: 

\begin{widetext}
\begin{eqnarray}
k_{B} T_{D} = \frac {\hbar\gamma_{_1}}{2} \cdot \frac{(k_{_1}^2 + k_{_2}^2)}{2(k_{_1} + k_{_2})} \cdot \left\{ \frac{k_{_1}\left|x_{_1}\right|}{1 + 4x_{_1}^2} + \frac{(k_{_1} + k_{_2})\left|x_{_2}\right| \gamma_{_1}/\gamma_{_2}}{1 + 4x_{_2}^2} \right\}^{-1}.
\label{eq:kBTD}
\end{eqnarray}
\end{widetext}

\bigskip

\noindent
We recognize the first term on Eq.~(\ref{eq:kBTD}), as the one-photon Doppler limit, $T_1 = \hbar \gamma_{_1}/2k_{B}$, which is $1.94~mK$, $831~\mu K$, $108~\mu K$ and $672~\mu K$, respectively for Magnesium, Calcium, Strontium and Ytterbium atoms. The detunings that minimize the two-photon Doppler temperature are $x_{_i} = - 1/2$ ($\delta\omega_{_1} = -\gamma_{_1}/2$, $\delta\omega_{_2} = - \gamma_{_2}/2$), giving the minimum value

\begin{eqnarray}
k_{B} T_{min} = \hbar\gamma_{_1} \cdot \frac{(k_{_1}^2 + k_{_2}^2)}{(k_{_1} + k_{_2})\cdot \left[k_{_1} + (k_{_1} + k_{_2}) \cdot (\gamma_{_1}/\gamma_{_2})\right]}.
\label{eq:kBTmin}
\end{eqnarray}

\bigskip
\noindent
Using the parameters of Table I we obtain the following temperatures: $T_{min}$ = $131~\mu K$, $123~\mu K$, $57~\mu K$ and $124~\mu K$, respectively for Magnesium, Calcium, Strontium and Ytterbium. These minimum values of temperature are very close for these elements because the linewidths of the second transition, $\gamma_{_2}$, are very similar (see Table I). This does not happen for the one-photon Doppler limit. The predicted ratio between the one- and two-photon Doppler temperatures will be then $\hbar\gamma_{_1}/(2k_{B}T_{min})$ = 14.8 ($Mg$), 6.8 ($Ca$), 1.9 ($Sr$) and 5.4 ($Yb$) (Table II). Equations (\ref{eq:kBTD}) and (\ref{eq:kBTmin}) are valid only for low laser intensities ($S_{1} << 1$, $S_{1} \cdot S_{2} << 1$). If we take into account higher intensities, the result is an increase of  temperatures, as in the case of ``one-photon optical molasses'' \cite{Lett89}.

\bigskip

\begin{center}
\begin{tabular}{p{1.05in} p{1.05in} p{1.05in}}
\hline
\hline 

~Element & $T_{min} (\mu K)$ ~~~& ~~~$T_1/T_{min}$ \\ \hline
$~~^{24}Mg$ &~~~~131 &~~~~14.8 \\ 
$~~^{40}Ca$ &~~~~123&~~~~~6.8 \\
$~~^{88}Sr$ &~~~~~57 &~~~~~1.9 \\ 
$~~^{174}Yb$ &~~~~124 &~~~~~5.4 \\ \hline
\hline
\end{tabular}
\end{center}

\begin{center}
Table II: Predicted parameters for the two-photon Doppler cooling.
\end{center}

\bigskip
\noindent
\textbf{B. Combined One- and Two-Photon Cooling Processes}

\bigskip

Since the one-photon $^1S_0$ - $^1P_1$ and two-photon $^1S_0$ - $^1S_0$ transitions occurs simultaneously, we should consider both cooling processes jointly. Depending on the detunings and intensities of the incident laser beams, one process can dominate the other. The case of ``one-photon optical molasses'' was discussed in several references, and results in the following coefficients \cite{Metcalf99}, \cite{Lett89}:

\begin{eqnarray} 
\alpha_{_1} = \frac{- 8 \hbar k_{_1}^2 S_{1}x_{_1}}{\left(1 + 2S_{1} + 4x_{_1}^2 \right)^2},
\label{eq:alpha1} 
\end{eqnarray} 

\begin{eqnarray}
D_1 = \frac{\hbar^2 k_{_1}^2 S_{1} \gamma_{_1}}{\left(1 + 2S_{1} + 4x_{_1}^2\right)},
\label{eq:D1}
\end{eqnarray}

\bigskip
\noindent
taking into account the saturation of the first transition involving the blue laser at frequency $\omega_{_1}$. The effective damping and diffusion coefficients that jointly take into account the one- and two-photon cooling processes are:

\begin{eqnarray}
\alpha_{eff} = \alpha_{_1} + \alpha_{_2},
\label{eq:alfa effective} 
\end{eqnarray} 

\begin{eqnarray}
D_{eff} = D_{_1} + D_{_2},
\label{eq:D effective} 
\end{eqnarray} 

\bigskip

\noindent
and the Doppler equilibrium temperature is obtained by $k_{B}T_{D} = D_{eff}/\alpha_{eff}$. Figure 3 presents this temperature for Calcium atoms, as a function of the infrared saturation parameter $S_2$, for several values of the first laser saturation parameter ($S_1 = 0.3$, $0.2$, $0.1$, and $0.01$). The laser detunings were assumed to be at the optimum values $\delta\omega_{_2} = -\gamma_{_2}/2$ and $\delta\omega_{_1} = - \gamma_{_1}/2$. Some features in Fig.~3 call our attention. The limit of low intensities of the infrared laser ($S_2 << 1$) results in a temperature expected on the basis of the two-level atom theory: $\hbar \gamma_{_1}/2k_{B} = 831~\mu K$ for Calcium. However, for low intensities of the blue laser ($S_1 << 1$) and with the increase of the $S_2$ parameter, the Doppler temperature is quite reduced, tending to the minimum value of $123~\mu K$ (dash dot-doted curve; see also Table II). This minimum  value of temperature is close to the Doppler limit that would be associated only with the second transition, that is, $\hbar \gamma_{_2}/2k_{B} = 115~\mu K$ for Calcium atoms. As already mentioned, this new Doppler limit is  $6.8$ times smaller than the one-photon Doppler limit of the blue transition. This happens due to the increase of the damping coefficient for the combined cooling processes. For the other elements of Table II this reduction factor varies considerably depending on the values of the linewidths $\gamma_{_1}$ and $\gamma_{_2}$.

\bigskip

\section{Efficiency of the cooling process and main loss mechanisms}

In this section, we discuss the main loss mechanisms that can limit the efficiency of the Doppler cooling scheme discussed here. It is important to know the population fraction in the excited states $\left|r \right\rangle$ and $\left|e \right\rangle$,  which can be lost by spontaneous emission into metastable levels that do not participate in the cooling process, or through photoionization induced by the incident laser beams.

Figure~4 shows some possible loss channels that can limit the cooling process, again using Calcium as an example. One of these channels is the spontaneous decay from the $(4s4p)^1P_1$ to the $(3d4p)^1D_2$ level, at a rate $\gamma_{PD} = 2180 s^{-1}$ \cite{Beverini89}. For Magnesium, the $^1D_2$ state is above the $^1P_1$ level, so this loss channel is not present. Another is the direct spontaneous decay from the $(4s5s) ^1S_0$ level to the $^{3}P_{1}$ level, at 553 nm, which occurs at a rate of $2440 (600) s^{-1}$ \cite{Curtis01}. A third loss channel is due to photoionization of the excited state $(4s5s) ^1S_0$, which can be connected to the $(3d6p) ^1P_1$ level above the ionization limit by a photon at 423 nm. The photoionization rate for this process is given by $\Gamma_{ion} = \sigma_{ion}(\omega_{1})$ x $ I(\omega_{1})/(\hbar \omega_{1})$ \cite{Chin84}, where $\sigma_{ion}(\omega_{1})$ is the photoionization cross-section, at the blue laser frequency. Although this cross-section has not been reported in the literature, we do not expect an expressive photoionization rate for Calcium, because this process can be considered far off-resonance. The branching ratio of the ionization process with respect to the spontaneous decay to the intermediate level $\left|r \right\rangle$, and from this to the ground state should be very low, because $\Gamma_{ion} << \gamma_{1}$, $\gamma_{2}$, for small values of intensity of the blue laser. This is not the case of Magnesium, where photoionization is an important loss channel.

The atomic populations of the excited states (Fig.1~(a)) can be calculated using the density matrix formalism, applied to a three-level atom interacting with two laser beams \cite{Wei98}. The optical Bloch equations (OBE's) for the density matrix elements, with the terms $\rho_{_{ij}} (i,j = g, r, e)$ varying slowly in the rotating-wave approximation (RWA), are given by:

\begin{widetext}
\begin{eqnarray}
\frac{\partial \rho_{ee}}{\partial t} & = & - \gamma_{_2} \rho_{ee} - i\cdot \frac{\Omega_{_2}}{2}\cdot (\rho_{er} - \rho_{re}), \nonumber
\label{eq:roee} 
\end{eqnarray} 

\begin{eqnarray}
\frac{\partial \rho_{rr}}{\partial t} & = & - \gamma_{_1} \rho_{rr} + \gamma_{_2} \rho_{ee}  - i\cdot \frac{\Omega_{_1}}{2} \cdot (\rho_{rg} - \rho_{gr}) + i\cdot \frac{\Omega_{_2}}{2} \cdot (\rho_{er} - \rho_{re}), \nonumber
\label{eq:rorr} 
\end{eqnarray} 

\begin{eqnarray}
\frac{\partial \rho_{gg}}{\partial t} & = & \gamma_{_1} \rho_{rr} + i\cdot \frac{\Omega_{_1}}{2} \cdot (\rho_{rg} - \rho_{gr}), \nonumber
\label{eq:rogg} 
\end{eqnarray} 

\begin{eqnarray}
\frac{\partial \rho_{er}}{\partial t} & = & - \left[ (\gamma_{_1} + \gamma_{_2})/2 - i (\delta\omega_{_2} - \delta\omega_{_1})\right]\cdot \rho_{er} - i \cdot \frac{\Omega_{_1}}{2} \cdot \rho_{eg} - i\cdot \frac{\Omega_{_2}}{2} \cdot (\rho_{ee} - \rho_{rr}), \nonumber
\label{eq:roer} 
\end{eqnarray} 

\begin{eqnarray}
\frac{\partial \rho_{eg}}{\partial t} & = & - \left[ \gamma_{_2}/2 - i \delta\omega_{_2}\right]\cdot \rho_{eg} - i\cdot \frac{\Omega_{_1}}{2} \cdot \rho_{er} + i\cdot \frac{\Omega_{_2}}{2} \cdot \rho_{rg}, \nonumber
\label{eq:roeg}
\end{eqnarray} 

\begin{eqnarray}
\frac{\partial \rho_{rg}}{\partial t} & = & - \left[ \gamma_{_1}/2 - i \delta\omega_{_1}\right]\cdot \rho_{rg} - i\cdot \frac{\Omega_{_1}}{2} \cdot (\rho_{rr} - \rho_{gg}) + i\cdot \frac{\Omega_{_2}}{2} \cdot \rho_{eg},
\label{eq:rorg} 
\end{eqnarray} 
\end{widetext}

\bigskip
\bigskip \noindent 
where atomic coherences obey the relation $\rho _{ij}^{*} = \rho _{ji}$. We have assumed in these equations that the linewidths of the incident laser beams are smaller than the half natural widths of the atomic transitions, that is, $\Delta \nu_1 < \gamma_{1}/2$, $\Delta \nu_2 < \gamma_{2}/2$. A linewidth of about a MegaHertz would satisfy this condition, and is tipical of, for example, a commercially available Ti:Sapphire laser.

The coupled OBE's in Eqs.(\ref{eq:rorg}) are numerically solved by integrating over the time $t$, via a Runge-Kutta fourth-order method, with the initial conditions: $\rho_{ij}(t = 0) = \delta_{ig}$. The populations $\rho_{ee}$ and $\rho_{rr}$ for Calcium atoms can be seen in Fig.5~(a) and (b) respectively, where the blue laser saturation parameter is $S_{1} = 0.1$, while the infrared saturation parameters are: $S_{2} = 1$ (doted), 3 (dashed) and 10 (solid line). We can observe in Fig.~5 that the increase of the infrared laser intensity results in an increase of the excited state populations, due to optical pumping. However, in the steady-state regime ($t \rightarrow \infty$), the percentage of excited atoms is low ($<10\%$), corresponding to a high fraction of atoms in the ground state.

An important parameter that characterizes the dynamics of the cooling process is the cooling time, $\tau_{cool}$. This time for the two-photon cooling can be calculated by the relationship between the kinetic energy and its loss rate:

\begin{eqnarray} 
\tau_{cool} = - \frac{E}{(dE/dt)_{cool}} = \frac{M}{2\alpha_{eff}}.
\label{eq:effective coefficients} 
\end{eqnarray} 

\bigskip

\noindent
Recalling equation (\ref{eq:alfa effective}), the last expression implies in a decrease of the cooling time, in comparison to one-photon cooling: $\tau_{cool} = \frac{M}{2\alpha_{1}} \cdot \frac{1}{1 + \alpha_{2}/\alpha_{1}} < \frac{M}{2\alpha_{1}}$. For Calcium we have $\tau_{cool} \approx ~ 2 \mu s$, for $\delta \omega_{_{i}} = - \gamma_{_i}/2$, $S_{1} = 0.1$ and $S_{2} = 3$. Since this time is much shorter than the storage time, dictated by optical pumping into the metastable $^1D_2$ state \cite{Andreas02}, two-photon cooling should be a fast and efficient process.

Another important parameter is the capture velocity, which is on the order of ~$\gamma_{_1}/k_{_1} = 14.7~ m/s$ for the one-photon cooling process in Calcium atoms  \cite{Metcalf99}. In general the average force written in equation (\ref{eq:Ftotal}) reduces to $F = -\alpha_{2} \cdot v/[1 + (v/v_{c})^2]$, and the capture velocity for the two-photon cooling is a little smaller, on the order of $v_{c} = (\gamma_{_1} + \gamma_{_2})/(2k_{_1} + 2k_{_2})$ for detunings $\left| \delta \omega_{_i} \right| = \gamma_{_i}/2$. For Calcium $v_{c} = 6~m/s$ is much larger than the mean velocity of the atoms already cooled in the first cooling process: $v_{rms} = \sqrt{\hbar\gamma_{_1}/(2M)} = 41.6~cm/s$. This fact ($v_{rms} << v_{c}$) implies that the initial number of atoms that were captured in the first cooling stage should be transferred completely into the second one, resulting in a significant number of colder atoms. In other words, the transfer efficiency from the first (one-photon) to the second (two-photon) stage should be basically $100\%$.

\section{Conclusions and Prospects}

We proposed a new possibility to reach microKelvin temperatures in laser cooling of alkaline-earth-metal and Ytterbium atoms, using a two-photon $^1S_0$ - $^1S_0$ transition. After discussing the implementation of this technique and the calculation of the transition rates, we analyzed the experimental cases of atomic interaction with a pair of copropagating laser beams and two-photon optical molasses. We considered excitation with laser beams at frequencies 
$\omega_{_1}$ and $\omega_{_2}$, in near resonance with the $^1P_1$ state, to enhance the two-photon scattering rate. Doppler limits of 131 $\mu K$ (Mg), 123 $\mu K$ (Ca), 57 $\mu K$ (Sr) and 124 $\mu K$ (Yb) have been found at optimum 
detunings and laser powers. Since the optimum detuning for the blue $^1S_0$ - $^1P_1$ transition is $\gamma_{_1}/2$, 
two-photon cooling can be simply applied as a second stage, just by adding the second near-infrared laser. This will 
bring all atoms to substantially lower temperatures (for example from 831 to 123 $\mu K$ for Calcium). With such 
$100\%$ efficiency, it should not be expected any loss of atoms in this second cooling stage. It is important to point 
out that two-photon cooling is considerably simpler than ``quenching'' cooling, requiring only one extra laser, with a 
linewidth of about a MegaHertz. The two-photon cooling scheme should be applicable to a $3D$ optical molasses or a 
magneto-optical trap, in this case taking advantage of the magnetic intermediate state $^1P_1$. This will be 
investigated in a future work. We also expect to make the experimental implementation of this technique with the 
addition of the 1034-nm laser in our 3D Calcium optical molasses and MOT \cite{Cavasso02}. Finally, the lower 
temperatures achieved by two-photon cooling can be also very useful for efficient loading of an optical dipole trap, 
which might be an important step towards the achievement of all-optical Bose-Einstein condensates \cite{Chapman01} for 
alkaline-earth-metal and Ytterbium atoms.

\begin{acknowledgments}
This work was supported by the Brazilian government agencies FAPESP (including its Optics and Photonics Center), CAPES, CNPq and FAEP-UNICAMP. FCC benefited from fruitful discussions with C. W. Oates, J. C. Bergquist, W. M. Itano, N. Beverini and A. Hemmerich.
\end{acknowledgments}


\begin{thebibliography}{50}

\bibitem{Dalibard89}
J. Dalibard and C. Cohen-Tannoudji, J. Opt. Soc. Am. B, \textbf{6}, 2023 (1989).
\bibitem{Weiss89}
D.S. Weiss, E. Riis, Y. Shevy, P.J. Ungar and S. Chu, J. Opt. Soc. Am. B, \textbf{6}, 2072 (1989).
\bibitem{Katori99}
H. Katori, T. Ido, Y. Isoya and M. Kuwata-Gonokami, Phys. Rev. Lett. \textbf{82}, 1116-1119 (1999).
\bibitem{Kuwamoto99}
T. Kuwamoto, K. Honda, Y. Takahashi and T. Yabuzaki, Phys. Rev. A \textbf{60}, R745-R748 (1999). 
\bibitem{Curtis01}
E.A. Curtis, C.W. Oates and L. Hollberg, Phys. Rev. A \textbf{64}, 31403-1-31403-4 (2001).
\bibitem{Binnewies01}
T. Binnewies, G. Wilpers, U. Sterr, F. Riehle, J. Helmcke, T.E. Mehlstaubler, E.M. Rasel and W. Ertmer, Phys. Rev. Lett. \textbf{87}, 123002 (2001).
\bibitem{Jim89}
F. Diedrich, J.C. Bergquist, W.M. Itano, D.J. Wineland, Phys. Rev. Lett. \textbf{62}, 403-406 (1989).
\bibitem{Andreas02}
J. Grunert and A. Hemmerich, Appl. Phys B \textbf{73}, 815-818 (2001); J. Grunert and A. Hemmerich,  Phys. Rev. A {\textbf 65}, 041401 (2002).
\bibitem{Kasevich92}
M. Kasevich and S. Chu, Phys. Rev. Lett \textbf{69}, 1741-1744 (1992).
\bibitem{Demtroder96}
W. Demtr\"{o}der, \textit{Laser Spectroscopy} (Springer, Berlin, 1996).
\bibitem{Cagnac73}
B. Cagnac, G. Grynberg and F. Biraben, Le Journal de Physique \textbf{34}, 845-858 (1973).
\bibitem{Loudon83}
R. Loudon, \textit{The Quantum Theory of Light} (Clarendon Press, London, 1983).
\bibitem{Bonin84}
K.D. Bonin and T.J. McIlrath, J. Opt. Soc. Am. B \textbf{1}, 52 (1984).
\bibitem{Metcalf99}
H.J. Metcalf and P. van der Straten, \textit{Laser cooling and trapping} (Springer-Verlag, New York, 1999).
\bibitem{Manoel02}
D.A. Manoel, R.L. Cavasso Filho, A. Scalabrin, D. Pereira and F.C. Cruz, Opt. Commun. \textbf{201}, 157-163 (2002).
\bibitem{Onisto02}
H.J. Onisto, R.L. Cavasso Filho, A. Scalabrin, D. Pereira and F.C. Cruz, Opt. Engineering \textbf{41}, 1122-1127 (2002).
\bibitem{Lett89}
P.D. Lett, W.D. Phillips, S.L. Rolston, C.E. Tanner, R.N. Watts and C.I. Westbrook, J. Opt. Soc. Am. B \textbf{6}, 2084-2107 (1989).
\bibitem{Beverini89}
N. Beverini, F. Giammanco, E. Maccioni, F. Strumia and G. Vissani, J. Opt. Soc. Am. B \textbf{6}, 2188-2193 (1989).
\bibitem{Chin84}
S.L. Chin and P. Lambropoulos, \textit{Multiphoton Ionization of Atoms}, (Academic Press, New York, 1984).
\bibitem{Wei98}
C. Wei, D. Suter, A. S. M. Windsor and N. B. Manson, Phys. Rev. A \textbf{58}, 2310 - 2318 (1998). 
\bibitem{Cavasso02}
R.L. Cavasso Filho, D.A. Manoel, D.R. Ortega, A. Scalabrin, D. Pereira and F.C. Cruz, Phys. Rev. A (submitted, 2002); R.L. Cavasso Filho, D.A. Manoel, D.R. Ortega, A. Scalabrin, D. Pereira and F.C. Cruz, in \textit{Proceedings of the $6^{th}$ Symposium on Frequency Standards and Metrology}, Scotland, 2001, edited by Patrick Gill (University of St Andrews, Scotland, 2001), p. 546.
\bibitem{Chapman01}
M. D. Barrett, J. A. Sauer and M. S. Chapman,~Phys. Rev. Lett. \textbf{87}, 010404 (2001).
\bibitem{Kurucz88}
R. L. Kurucz,~Trans.~IAU,~XXB,~M. McNally,~ed. Dordrecht: Kluwer,~168-172 (1988).
\end{thebibliography}
\end{document}